\documentclass[prl,aps,twocolumn,preprintnumbers,amsmath,amssymb,nofootinbib,superscriptaddress,notitlepage]{revtex4-1}

\usepackage{epsfig}
\usepackage[utf8]{inputenc}
\usepackage{color}
\usepackage{epstopdf}
\usepackage[colorlinks=true,
linkcolor=blue,
breaklinks=true,
urlcolor=blue,
citecolor=blue]{hyperref}

\newcommand{\uestc}{\affiliation{School of Physics, University of Electronic Science and
Technology of China, Chengdu 610054, China}} 

\newcommand{\itp}{\affiliation{CAS Key Laboratory of Theoretical Physics, Institute of Theoretical Physics,\\ Chinese Academy of Sciences, Beijing 100190, China}}

\newcommand{\ucas}{\affiliation{University of Chinese Academy of Sciences, Beijing 100049, China}}

\newcommand{\imp}{\affiliation{Institute of Modern Physics, Chinese Academy of Sciences, Lanzhou 730000, China}}

\begin{document}

\title{Strange molecular partners of the $Z_c(3900)$ and $Z_c(4020)$}

\author{Zhi Yang}\email{zhiyang@uestc.edu.cn}
\uestc

\author{Xu Cao}\email{caoxu@impcas.ac.cn}
\imp
\ucas

\author{Feng-Kun Guo}\email{fkguo@itp.ac.cn}
\itp
\ucas

\author{Juan Nieves}\email{jmnieves@ific.uv.es}
\affiliation{Instituto de F\'isica Corpuscular (centro mixto CSIC-UV),
Institutos de Investigaci\'on de Paterna, Apartado 22085, 46071, Valencia, Spain
}

\author{Manuel Pavon Valderrama}\email{mpavon@buaa.edu.cn}
\affiliation{School of Physics and Nuclear Energy Engineering, \\
International Research Center for Nuclei and Particles in the Cosmos and \\
Beijing Key Laboratory of Advanced Nuclear Materials and Physics, \\
Beihang University, Beijing 100191, China} 

\date{\today}

\begin{abstract} 
  \rule{0ex}{3ex}
  Quantum Chromodynamics presents a series of exact and approximate symmetries which can be exploited  to predict new hadrons from previously known ones.
  The $Z_c(3900)$ and $Z_c(4020)$, which have been theorized to be
  isovector $D^* \bar{D}$ and $D^* \bar{D}^*$ molecules [$I^G(J^{PC}) = 1^-(1^{+-})$], are no exception.
  Here we argue that from SU(3)-flavor symmetry, we should expect
  the existence of strange partners of the $Z_c$'s with hadronic molecular
  configurations $D^* \bar{D}_s$-$D \bar{D}_s^*$ and $D^* \bar{D}^{*}_s$
  (or, equivalently, quark content $c\bar{c} s \bar{q}$,
  with $q = u, d$).
  The quantum numbers of these $Z_{cs}$ and $Z_{cs}^*$ {structures} would be
  $I(J^P)$ = $\frac{1}{2}(1^+)$.
  The predicted masses of these partners depend on the details of the theoretical
  scheme used, but they should be around the $D^* \bar{D}_s$-$D \bar{D}_s^*$ and $D^* \bar{D}^{*}_s$ thresholds, respectively. { Moreover, any of these states could be either a virtual pole or a  resonance.}
  We show that, together with a possible triangle singularity contribution, such a picture nicely agrees with the very recent BESIII data of the $e^+e^-\to K^+(D_s^-D^{*0}+D_s^{*-}D^0)$.
\end{abstract}

\maketitle

{\it{Introduction.}}--- Unsuccessful searches  for charged charmonium-like states ($Z^{(*)}_{cs}$) with hidden-charm and open-strange channels in $e^+ e^- \to K^+ K^- J/\psi$ were reported by Belle~\cite{Shen:2014gdm,Yuan:2007bt} and BESIII~\cite{Ablikim:2018epj}. Recently, however, the BESIII Collaboration observed~\cite{Ablikim:2020hsk} a new strange, hidden-charm {state} with a mass and width of
\begin{eqnarray}
  M(Z_{cs}) = 3982.5^{+2.8}_{-3.3} \,{\rm MeV} , ~~
  \Gamma(Z_{cs}) = 12.8^{+6.1}_{-5.3} \,{\rm MeV} .~~~\label{eq:zcs_bes}
\end{eqnarray}
This raises the question of what its nature is.
Its closeness to the two-meson $D^* \bar{D}_s$-$D \bar{D}_s^*$ thresholds immediately
suggests the possibility that it might be a hadronic molecule. 
Theoretical predictions of $Z_{cs}$ states have been made in different models~\cite{Lee:2008uy,Voloshin:2019ilw,Ferretti:2020ewe,Chen:2013wca} but in general lie considerably above the $D\bar D^*_s/ D^*\bar D_s$ and $D^*\bar D^*_s/ \bar D^*D_s^*$ thresholds (with few exceptions~\cite{Lee:2008uy,Chen:2013wca}).

To evaluate the reliability of the molecular hypothesis, we compare the new $Z_{cs}^{(*)}$ {state} with other known molecular candidates.
Of particular relevance are 
the $Z_c(3900)$ and $Z_c(4020)$~\cite{Ablikim:2013mio,Liu:2013dau, Ablikim:2013wzq} ($Z_c$ and $Z_c^*$ from now on),
two charged hidden-charm states which proximity to the
$D\bar{D}^*$ and $D^* \bar{D}^*$ thresholds, respectively~\cite{Zyla:2020zbs},
suggested their molecular nature.
A decade ago the Belle Collaboration discovered
the $Z_b(10610)$ and $Z_b(10650)$ ($Z_b$ and $Z_b^*$), a pair of charged
hidden-bottom {states} with $I^G(J^{PC}) = 1^-(1^{+-})$
and masses
also very close to the $B\bar{B}^*$ and $B^* \bar{B}^*$ thresholds~\cite{Belle:2011aa},
which raised the question whether the $Z_b$'s were indeed
bound states of the bottom mesons.

{\it{Symmetries.}}--- The fact that the $Z_c$'s and $Z_b$'s come in pairs is naturally
explained within the molecular picture from heavy-quark spin symmetry (HQSS)
considerations. This happens when the off-diagonal piece between the $D\bar D^*(B\bar B^*)$ and  $D^*\bar D^*(B^*\bar B^*)$ is neglected. Such a phenomenon was called 
``light-quark spin symmetry" by Voloshin~\cite{Voloshin:2016cgm}, where experimental observations are consistent with it.
In addition, the existence of  hidden-charm states can be deduced
from the hidden-bottom ones and the heavy-flavor symmetry for the potential~\cite{Guo:2013sya},
though whether the relation between the hidden charm and bottom sectors
is of a phenomenological or of a systematic nature has recently been challenged~\cite{Baru:2018qkb}.
The question we would like to address in this work is whether
the new $Z_{cs}$ {structure} is related to the $Z_c$ and $Z_c^*$. 
{We use the nomenclature ``structure", because the signatures reported by BESIII might also be due to a virtual state ---pole in unphysical Riemann sheet close to the $D^* \bar{D}_s$-$D \bar{D}_s^*$ threshold--- instead of a resonance, when lower channels are neglected.} 

Besides heavy quarks, the $Z_c^{(*)}$'s also contain light-quarks
and are constrained by SU(2)-isospin and SU(3)-flavor symmetries~\cite{HidalgoDuque:2012pq,Peng:2019wys}.
If we generically denote the $D^{(*)}$ and $D_s^{(*)}$ mesons as $D_a^{(*)}$, 
with $a$ indicating flavor, 
then $D_a^{(*)}$ belongs to
the $\bar{3}$ representation of SU(3).
Thus
the $D_a^{(*)} \bar{D}^{(*)b}$ system contains singlet and octet irreducible representations: $3 \otimes \bar{3} = 1 \oplus 8$.
The flavor structure of the $D_a^{(*)}\bar{D}^{(*)b}$ potential, in a given $J^P$ sector with definite $C$-parity for the flavor neutral states,  
is
\begin{eqnarray}
  V = \lambda_S \, V^{(S)} + \lambda_O \, V^{(O)} \, ,
\end{eqnarray}
with $V^{(S)}$ and $V^{(O)}$ the singlet and octet parts of
the potential.
We obtain for the isoscalar and isovector $D^{(*)} \bar{D}^{(*)}$ systems, 
\begin{eqnarray}
  V(D^{(*)}\bar{D}^{(*)}, I=0) &=& \frac{2}{3} V^{(S)} + \frac{1}{3} V^{(O)} \, , \\
  V(D^{(*)}\bar{D}^{(*)}, I=1) &=& V^{(O)} \, , \label{eq:isovec}
\end{eqnarray}
while for $D^{(*)} \bar{D}_s^{(*)}$ and $D_s^{(*)} \bar{D}_s^{(*)}$, we find
\begin{eqnarray}
  V\big(D^{(*)}\bar{D}_s^{(*)}\big) = V^{(O)} , 
\ V(D_s^{(*)}\bar{D}_s^{(*)}) = \frac{V^{(S)}}{3}  + \frac{2 V^{(O)}}{3}  .~~~\label{eq:zcs} 
\end{eqnarray}
Of course, SU(3)-flavor symmetry is not exact, and we  might expect
these relations to be violated at about % a $20\%$ level. 
the $20\%$ level. 

Note that in the SU(3) limit, the isovector $D^{(*)}\bar{D}^{(*)}$ and the $D^{(*)}\bar{D}_s^{(*)}$ potentials  (Eqs.~\eqref{eq:isovec} and \eqref{eq:zcs}) are equal, which implies the existence of $Z_{cs}$ and  $Z_{cs}^*$ partners with  $I(J^P)= 1/2(1^+)$ of the $Z_c(3900)$ and $Z_c^*(4020)$. {
We point out that, when using a contact-range interaction, the microscopic exchange is not resolved.
It is sensible to think that the mass difference between the pion and $\eta$ mesons will generate sizable SU(3)-breaking effects.
However, Ref.~\cite{Aceti:2014uea} argues that
the exchange of light mesons in the octet sector is Okubo-Zweig-Iizuka suppressed in the SU(3) limit, 
as easily illustrated with $D^0 D^{*-} [c\bar u\, d \bar c]$ (third component of isospin $I_3=-1$)
where the two charmed mesons do not have the same light-quark flavor. As a consequence, 
the strength of the light-meson exchange potential is small and SU(3)-breaking corrections are not expected to be large.
On the other hand,  
Ref.~\cite{Aceti:2014uea} also finds that,
although two-pion exchange is really small, 
$J/\psi$-exchange plays a major role.  Moreover, 
Ref.~\cite{Dong:2021juy} argues that
since the exchanged $c\bar c$ state is highly off-shell, there are no  mass hierarchy arguments to select just the ground state $J/\psi$. 
It could well be the case that the entire series of $\psi$ states contribute to generate a strong potential, and the formation of the $Z_c$ and $Z_{cs}$ would be mainly due to this very short-range interaction. This type of potential is light-flavor independent and thus expected to be SU(3)-symmetric:
breaking corrections of the order of 20\% would lead to overly conservative estimates.}

Predictions for the $Z_{cs}$ and  $Z_{cs}^*$  masses depend
on the details of the potential, but we can make a first approximation
by assuming that the charmed mesons are massive enough as to ignore
the kinetic energy term in the Schr\"odinger equation,
\begin{eqnarray}
  H = T_{\rm kin} + V \approx V \quad \mbox{(for $m_{D_a}, m_{D^*_a} \to \infty$)} \, ,
\end{eqnarray}
in which case the binding energy of the molecule is given by
the matrix element of the potential,
i.e. $E_B \simeq \langle V \rangle$.
Within this approximation 
\begin{eqnarray}
  M(Z_c^*) - 2 m_{D^*} \simeq  M(Z_{cs}^*) - (m_{D^*}+m_{D^*_s}) \, , \label{eq:masses}
\end{eqnarray}
and an analogous relation for the $Z_c$ and $Z_{cs}$ states.
This will translate into predicted masses of around $3.99$~GeV and $4.13$~GeV for the $Z_{cs}$ and $Z_{cs}^*$ {states}, respectively.

{\it{EFT description.}}--- 
To get more accurate predictions,
we propose a concrete form for the potential.
The most general way to derive the interaction is
from an effective field theory (EFT).
For the $Z_c$'s the lowest order potential is usually a contact-range interaction without derivatives~\cite{Mehen:2011yh,Valderrama:2012jv}
\begin{eqnarray}
  V^{(O)}_{\rm virtual} = C^{(O)} \, . \label{eq:pot-virtual}
\end{eqnarray}
This interaction is able to generate a pole below its respective two-meson
threshold (a bound or virtual state), but not above threshold.
This is what might be happening for the $Z_c$ and $Z_c^*$,
which Breit-Wigner (BW) masses are around 
$11$ and $7\,{\rm MeV}$ above their respective thresholds,
although it is perfectly possible that the physical poles could very
well be below threshold~\cite{Albaladejo:2015lob}.
Alternatively we can use a different EFT suited for a resonant state, where
the contact-range potential reads
\begin{eqnarray}
  V^{(O)}_{\rm res} = C^{(O)} + 2 D^{(O)}\,k^2\, , \label{eq:pot-res}
\end{eqnarray}
with $k$ the c.m. momentum of
the two mesons.

In addition, the  potentials have to be regularized, included in a dynamical equation to obtain the poles and then renormalized.
We use a Gaussian regulator,
\begin{eqnarray}
  \langle p' | V^{(O)}_{\Lambda} | p \rangle = V^{(O)}\,
  g(\frac{p'}{\Lambda})\,g(\frac{p}{\Lambda}) \, ,
\end{eqnarray}
with $g(x) = e^{-x^2}$ and $V^{(O)}$ the unregularized potential of
Eqs.~(\ref{eq:pot-virtual}) or (\ref{eq:pot-res}), where the low-energy constants (LECs)
now depend on the cutoff, i.e., $C^{(O)} = C^{(O)}(\Lambda)$
and $D^{(O)} = D^{(O)}(\Lambda)$.

For a separable potential the Lippmann-Schwinger equation, $T = V + V G_0 T$, admits the ansatz
\begin{eqnarray}
  \langle p' | T(E_{\rm cm}) | p \rangle = \tau(E_{\rm cm})\,g(\frac{p'}{\Lambda})\,g(\frac{p}{\Lambda})
\end{eqnarray}
with $\tau(E_{\rm cm})$ given by
\begin{eqnarray}
\frac{1}{\tau(E_{\rm cm})} &=& {\frac{1}{C^{(O)} + 2 D^{(O)} k^2} - I_0(E_{\rm cm}; \Lambda)} \, ,
\label{eq:tau}\\
  I_0(E_{\rm cm}; \Lambda) &=& 
  \int \frac{d^3 q}{(2 \pi)^3}\,
  \frac{g^2(\frac{q}{\Lambda})}{E_{\rm cm} - M_{\rm th}-\frac{\vec{q}^{\,2}}{2\mu}+ i\epsilon} \, ,
  \label{eq:loop}
\end{eqnarray}
where $E_{\rm cm}$ is the c.m. energy of the two-body system, $M_{\rm th}= m_1 + m_2$ and $\mu=m_1m_2/(m_1+m_2)$, with $m_1$, $m_2$ the meson masses.
The $Z_c^{(*)}$ and $Z_{cs}^{(*)}$ states correspond to poles of the $T$-matrix, i.e.
to ${1}/{\tau(E_{\rm pole})} = 0$, in appropriate Riemann sheets.
Finally, we notice that the  $D \bar{D}_s^*$ and $D^* \bar{D}_s$ thresholds are separated by only $2.5\,{\rm MeV}$, which makes reasonable to simply approximate them by their average.

{\it{Determination of the LECs (previous data)}}---
For the determination of the couplings, first we fit to the location of the $Z_c$ and $Z_c^*$ poles as extracted in Ref.~\cite{Albaladejo:2015lob}.
In the {\it constant-contact} EFT without the $D^{(O)}$ term, we obtain
for $\Lambda = 0.5 (1.0)\,{\rm GeV}$
\begin{eqnarray}
  C^{(O)}(\Lambda) &=& -0.29^{+0.15}_{-0.32} \,
  \left(-0.28^{+0.08}_{-0.39}\right)\, {\rm fm}^2 \, ,
  \label{eq:co_zc}
\end{eqnarray}
while for the {\it resonant} EFT with both the $C^{(O)}$ and $D^{(O)}$ terms, we find
\begin{eqnarray}
  C^{(O)}(\Lambda) &=&
  -0.06^{+0.24}_{-0.16} \, \left(-0.22^{+0.10}_{-0.06}\right)\, {\rm fm}^2 \, , \notag
  \\
  D^{(O)}(\Lambda) &=& -0.31^{+0.10}_{-0.17}\,
  \left(-0.09^{+0.03}_{-0.07}\right)\,
  {\rm fm}^4 \, .
  \label{eq:cd_zc}
\end{eqnarray}
The predictions for the spectrum with this method are summarized in the upper half of Table \ref{tab:predictions}.
%

%------------------------------------------------------------------------
\begin{table*}[tb]
    \caption{\label{tab:predictions}
      Pole positions (MeV units) of the $D^{(*)a} \bar{D}_a^{(*)}$ molecules with $J^P = 1^{+}$. Non-strange states have $I^G=1^-$ and negative charge-conjugation quantum numbers (for the neutral ones).
      For the numerical calculations, we have used $m_D = 1867.2\,{\rm MeV}$,
      $m_{D^*} = 2008.6\,{\rm MeV}$, $m_{D_s} = 1968.3\,{\rm MeV}$ and 
      $m_{D_s^*} = 2112.2\,{\rm MeV}$. 
      We show results from both the constant-contact and  resonant EFTs introduced in Eqs.~(\ref{eq:pot-virtual})
      and ~(\ref{eq:pot-res}).
      The LECs are determined in two ways, either by reproducing the $Z_c$ pole obtained in Ref.~\cite{Albaladejo:2015lob} (top first two sets of pole positions), or by directly fitting to the BESIII data of $e^+e^-\to K^+(D_s^-D^{*0}+D_s^{*-}D^0)$~\cite{Ablikim:2020hsk} (bottom two sets). 
      If there is no imaginary part, the pole corresponds to a virtual state.
      For fit II we have propagated the errors in quadrature, which might result in an overestimation of the uncertainties for the masses since we are not considering correlations. We note that this fit allows both for resonant (R) and virtual state (V) solutions.
      For comparison, the $Z_c^{(*)}$ masses from Ref.~\cite{Zyla:2020zbs} and the $Z_{cs}$ pole position from BESIII~\cite{Ablikim:2020hsk} are also shown.
    }
    \begin{ruledtabular}
  \begin{tabular}{l c c c c c}
         Potential & States & Thresholds & Masses ($\Lambda=0.5$ GeV) &
         Masses ($\Lambda=1$ GeV) & Experiment~\cite{Ablikim:2020hsk,Zyla:2020zbs} 
         \\\hline
         $V^{(O)}_{\rm virtual}$ & $\frac1{\sqrt{2}}(D\bar D^*-D^*\bar D)$ &
         3875.8 &  Input~\cite{Albaladejo:2015lob} &
         Input~\cite{Albaladejo:2015lob} &
         {$3888.4 \pm 2.5 - i(14.2 \pm 1.3)$}  \\{}
         \!\![Eq.~(\ref{eq:co_zc})] & $D^* \bar{D^*}$ &
         4017.2 &
         $3988^{+21}_{-27} $ &
         $3978^{+25}_{-36}$ &
         $4024.1 \pm 1.9 - i(6.5\pm2.5)$ \\
         & $D \bar{D}^*_s/D^* \bar{D}_s$ &
         3979.4/3976.9 &
         $3948^{+22}_{-27} $ &
         $3937^{+25}_{-36}$ &
         \\
         & $D^* \bar{D}^*_s$ &
         4120.8 &
         $4092^{+21}_{-26}$ &
         $4083^{+24}_{-35}$ & \\
         \hline
         $V^{(O)}_{\rm res}$ & $\frac1{\sqrt{2}}(D\bar D^*-D^*\bar D)$ &
         3875.8 &
         Input~\cite{Albaladejo:2015lob} &
         Input~\cite{Albaladejo:2015lob} &
         {$3888.4 \pm 2.5 - i(14.2 \pm 1.3)$}\\{}
         \!\![Eq.~(\ref{eq:cd_zc})] & $D^* \bar{D^*}$ &
         4017.2 &
         $4025 \pm 4 - i (21 \pm 7)$ &
         $4035 \pm 6 - i (29 \pm 13)$ &
          $4024.1 \pm 1.9 - i(6.5\pm2.5)$ \\
         & $D \bar{D}^*_s/D^* \bar{D}_s$ &
         3979.4/3976.9 &
         $3986 \pm 4 - i (22 \pm 7)$ &
         $3996 \pm 6 - i (30 \pm 13)$ & 
         $3982.5^{+2.8}_{-3.3} - i (6.4^{+3.0}_{-2.7})$
         \\
         & $D^* \bar{D}^*_s$ &
         4120.8 &
         $4129 \pm 4 - i (21 \pm 7)$ &
         $4138 \pm 6 - i (28 \pm 12)$ &
         \\
      \hline
        \hline 
         $V^{(O)}_{\rm virtual}$ & $\frac1{\sqrt{2}}(D\bar D^*-D^*\bar D)$ &
         3875.8 & $3871_{-3}^{+2}$ &
         $3867^{+4}_{-7}$  &
         {$3888.4 \pm 2.5 - i(14.2 \pm 1.3)$} \\
         \!\![fit I, Eq.~\eqref{eq:fit-virtual}] & $D^* \bar{D^*}$ &
         4017.2 &
         $4014_{-3}^{+2}$  &
         $4012^{+3}_{-6}$  &
         $4024.1 \pm 1.9 - i(6.5\pm2.5)$\\
         & $D \bar{D}^*_s$/$D^* \bar{D}_s$ &
         3979.4/3976.9 &
         $3974_{-3}^{+2}$ &
         $3971^{+3}_{-6}$  &
         \\
         & $D^* \bar{D}^*_s$ &
         4120.8 &
         $4117_{-5}^{+3}$   &
         $4115^{+3}_{-6}$  & 
         \\\hline
         $V^{(O)}_{\rm res}$ & $\frac1{\sqrt{2}}(D\bar D^*-D^*\bar D)$ &
         3875.8 &
         $3861^{+15}_{-5} - i 6^{+11}_{-6}$ (R/V) &
         $3861^{+16}_{-29} - i 0^{+34}_{-0}$ (R/V)
         &
         {$3888.4 \pm 2.5 - i(14.2 \pm 1.3)$} \\{}
         \!\![fit II, Eq.~\eqref{eq:fit-res-d}] & $D^* \bar{D^*}$ &
         4017.2 &
         $4004^{+14}_{-6} - i 0^{+15}_{-0}$ (R/V) &
         $4006^{+11}_{-32} - i 0^{+30}_{-0}$ (R/V) &
         $4024.1 \pm 1.9 - i(6.5\pm2.5)$ \\
         & $D \bar{D}^*_s$/$D^* \bar{D}_s$ &
         3979.4/3976.9 &
         $3963^{+15}_{-5} - i 3^{+17}_{-3}$ (R/V) &
         $3966^{+13}_{-31} - i 0^{+31}_{-0}$ (R/V) &
         $3982.5^{+2.8}_{-3.3} - i (6.4^{+3.0}_{-2.7})$ \\
         & $D^* \bar{D}^*_s$ &
         4120.8 &
         $4110^{+11}_{-5} - i 0^{+15}_{-0}$ (R/V) &
         $4111^{+10}_{-23} - i 0^{+28}_{-0}$ (R/V)  &
     \end{tabular}
  \end{ruledtabular}
  \end{table*}
%------------------------------------------------------------------------

%-----------------------
\begin{figure}[t]
    \centering
    \includegraphics[width=\linewidth]{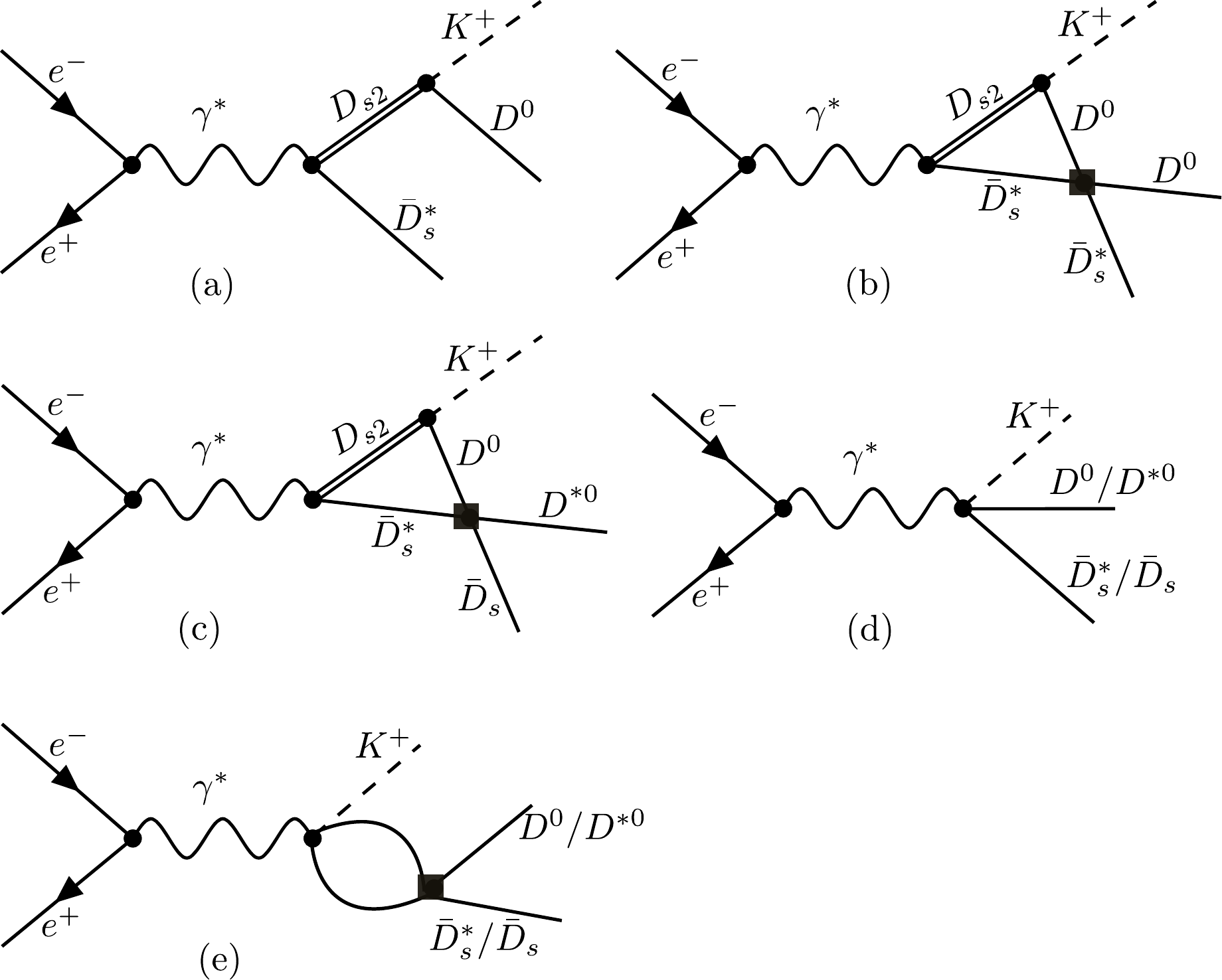}
    \caption{Feynman diagrams for the production mechanisms considered in this work:  (a) and (b) for the $K^+D_s^*\bar D^0$; (c) for the $K^+D_s\bar D^{*0}$; (d) and (e) for both final states. The filled squares denote the $T$-matrix elements which include the effects of the generated $Z_{cs}$ state.}
    \label{fig:feyn}
\end{figure}
\begin{figure}[tb]
    \centering
    \includegraphics[width=\linewidth]{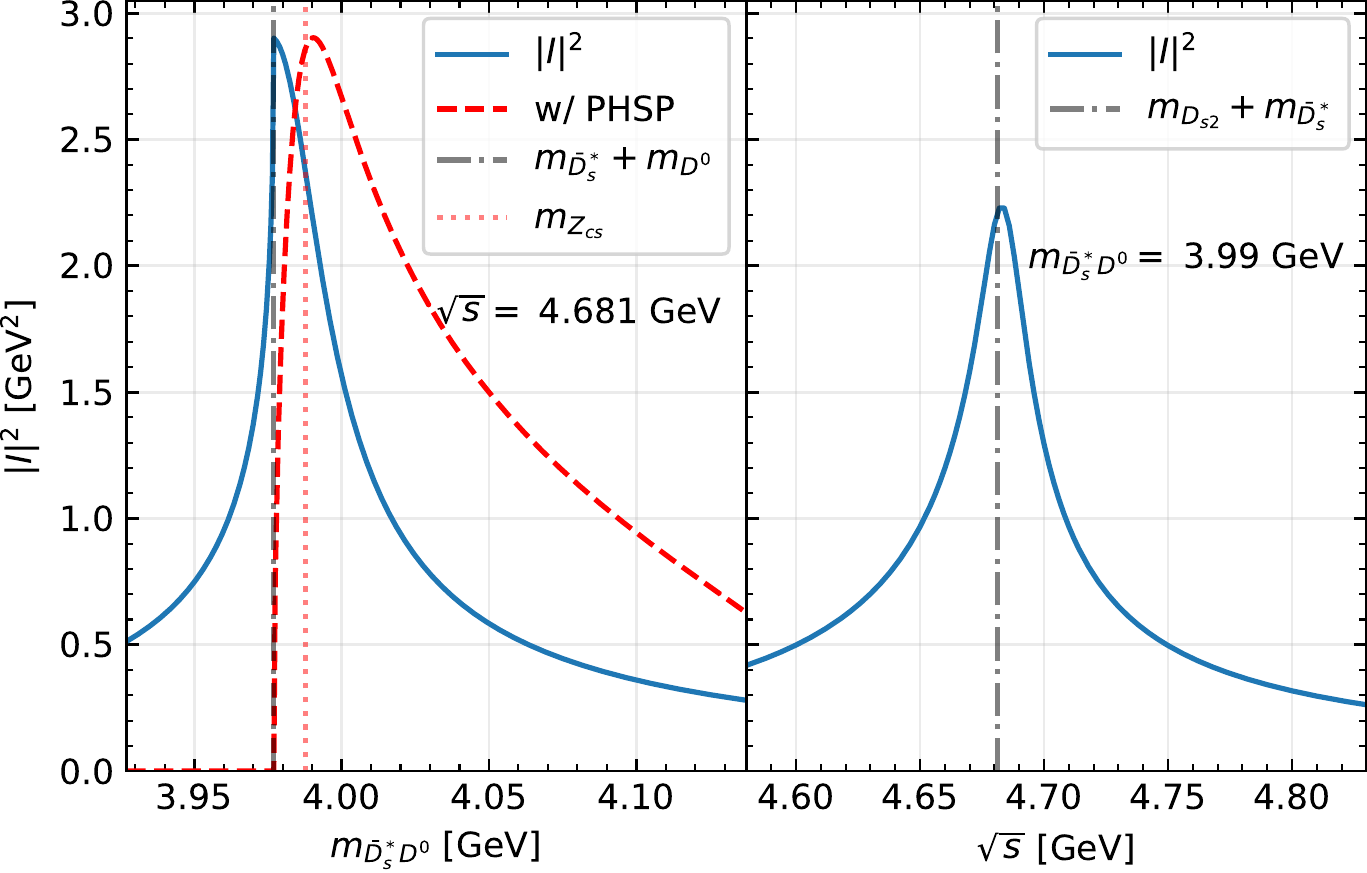}
    \caption{Absolute squared value  of the scalar triangle loop integral, $|I|^2$, with the $D_{s2}\bar D_s^* D^0$ intermediate state shown as Fig.~\ref{fig:feyn}(b). Left: dependence on the $\bar D_s^* D^0$ invariant mass for $\sqrt{s}=4.68$~GeV, where we also show $|I|^2$ convoluted with the phase space, with the maximum normalized to that of $|I|^2$; right: dependence on $\sqrt{s}$ with $m_{\bar D_s^* D^0}= 3.99$~GeV.}
    \label{fig:zs}
\end{figure}
%-----------------------

{\it{Analysis of the new data.}}--- The new measurements of the $e^{+}e^{-} \to K^{+}(D_s^* \bar{D}^0 + D_s \bar{D}^{*0})$ data with the $e^+e^-$ c.m. energy $\sqrt{s}=4.628$-$4.698\,{\rm GeV}$ allow us to determine the LECs from an independent source.
For this process one readily notices that there are triangle diagrams\footnote{{ Note the $D_{s1}(2536)$ can also give rise to a triangle singularity with a final $K D^*\bar D_s$ state,   when it is produced in association with the $\bar D_s$ in the    $e^+e^-$ annihilation. However, the $D_{s1}(2536)\bar D_s$ threshold is more than 100 MeV below the energy region of the BESIII measurements, and the effects of this mechanism are expected to be smaller than those derived from the $D_{s2}(2573)$. }} shown as diagrams (b) and (c) in Fig.~\ref{fig:feyn}. The triangle diagrams are special in the sense that they possess a triangle singularity~\cite{Landau:1959fi} when $\sqrt{s}\simeq m_{D_{s2}}+m_{\bar D^*}=4.681$~GeV. A triangle singularity  happens when all the intermediate particles in a triangle diagram are on their mass shell and move col-linearly so that the whole process may be regarded as a classical process in the space-time~\cite{Coleman:1965xm} (for a review, see Ref.~\cite{Guo:2019twa}).
On the one hand, triangle singularities produce peaks mimicking the resonance behavior; one the other, % hand, 
they can enhance the production of near-threshold hadronic molecules~\cite{Guo:2017jvc,Guo:2019twa}.
The importance of the $D_1(2420)\bar D D^*$ triangle diagrams for the $Z_c$ structures is discussed in Refs.~\cite{Wang:2013cya,Wang:2013hga,Liu:2014spa,Albaladejo:2015lob,Pilloni:2016obd,Guo:2020oqk}.
Here, one finds that the production of the $Z_{cs}$ can be facilitated by the $D_{s2}\bar D^* D^0$ triangle diagrams shown in Fig.~\ref{fig:feyn}.
To see this clearly, we show the absolute value squared of the corresponding scalar triangle loop integral, $|I|^2$, in Fig.~\ref{fig:zs}.
For  $\sqrt{s}=4.68$~GeV, $|I|^2$ convoluted with the three-body phase space for $e^+e^-\to K^+\bar D_s^* D^0$ has a clear peak around 3.99~GeV. Thus, such an effect needs to be taken into account when extracting information of $Z_{cs}$ from the data.
The expression for $I$ can be found in Refs.~\cite{Guo:2010ak,Guo:2017jvc,Guo:2019twa}.
To account for the finite width, 16.9~MeV, of the $D_{s2}(2573)$, we use a complex value $(2569.1-i8.5)$~MeV~\cite{Zyla:2020zbs} as its mass.

The amplitudes corresponding to the diagrams shown in Fig.~\ref{fig:feyn} can be easily worked out within the non-relativistic approximation for all the charmed mesons (for explicit expressions, we refer to the Supplemental), which can be used to fit the invariant mass spectra measured by BESIII.

The BESIII data were collected at five  c.m. $e^{+}e^{-}$ energies ranging from 4.628 to 4.698 
GeV. 
%-------------------------------------------------------
\begin{figure}[tb]
    \centering
    \includegraphics[width=\linewidth]{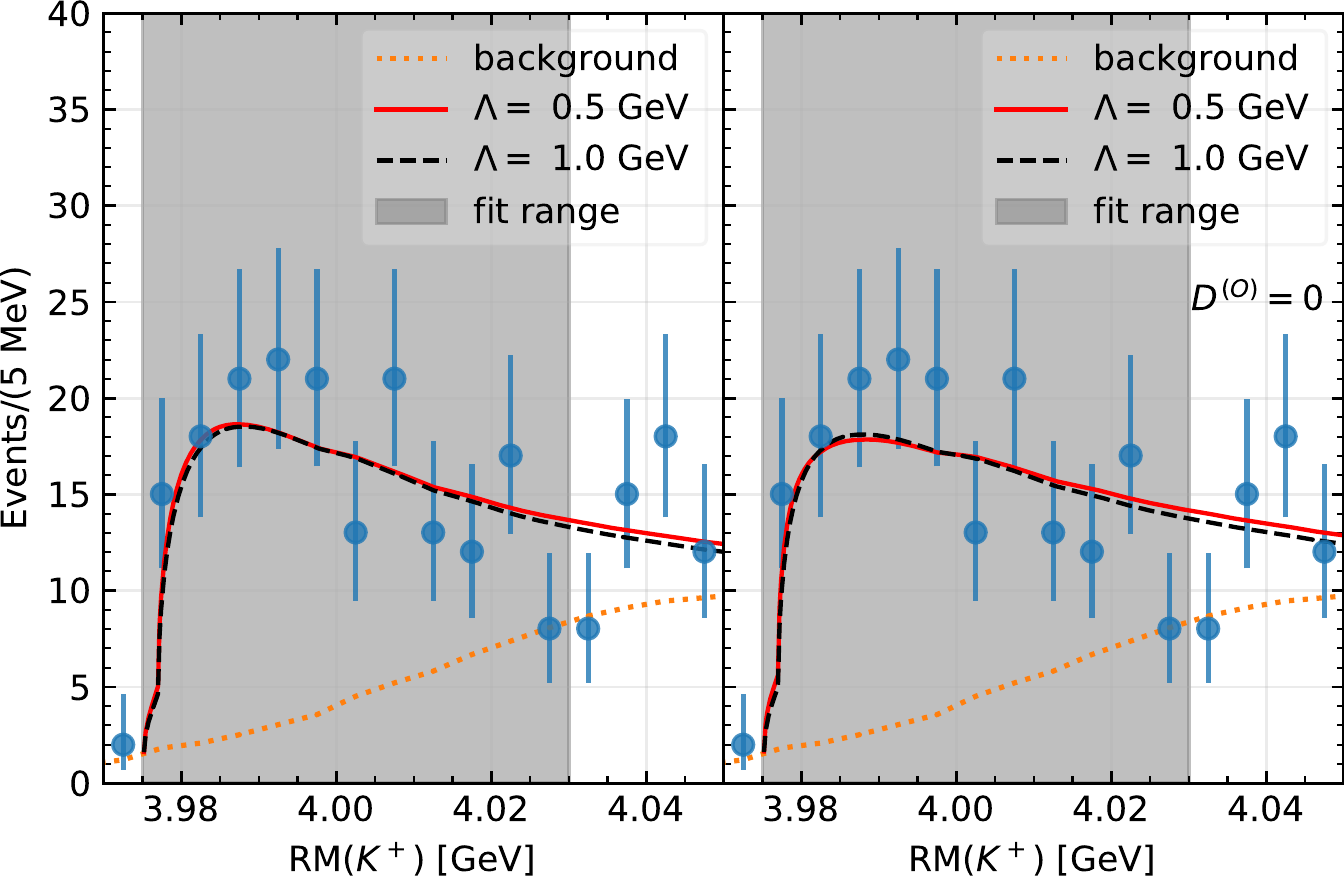}
    \caption {Fits to the BESIII event-spectrum for $\sqrt{s}=4.681$~GeV \cite{Ablikim:2020hsk}, as a function of ${\rm RM}(K^+)=\sqrt{(p_{e^-}+p_{e^+}-p_{K^+})^2}$. Left: best-fit results with both $C^{(O)}$ and $D^{(O)}$ taken as free parameters.  Right: best-fit results with $D^{(O)}=0$. The background contribution, in both plots, is taken as the { combinatorial} background in the BESIII analysis.}
    \label{fig:fit_b}
\end{figure}
\begin{figure}[tb]
    \centering
    \includegraphics[width=\linewidth]{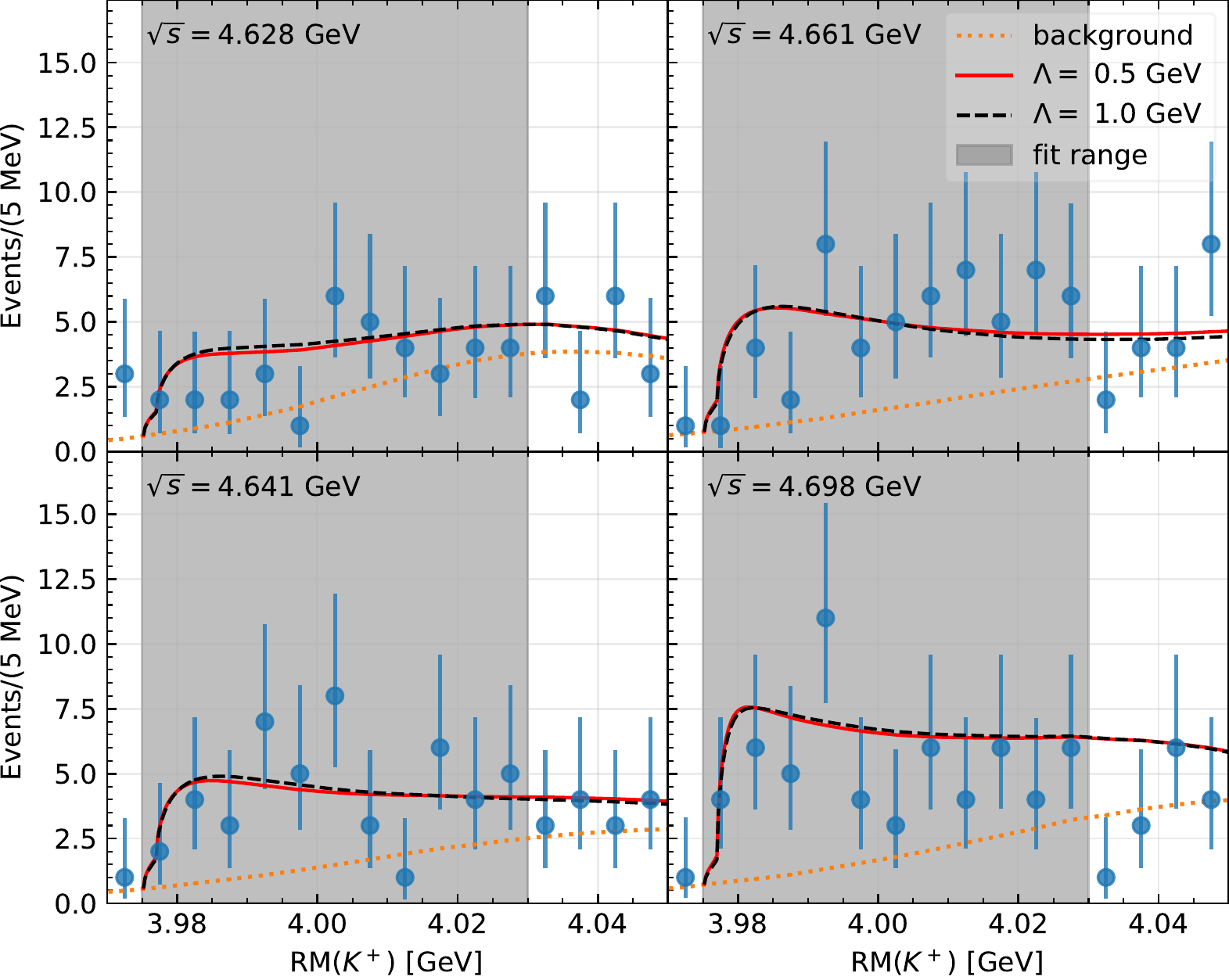}
    \caption{Best-fit results, with both $C^{(O)}$ and $D^{(O)}$ taken as free parameters, for the event-spectra reported by BESIII for another four c.m. $e^{+}e^{-}$ energies \cite{Ablikim:2020hsk}.}
    \label{fig:fit_b_4E}
\end{figure}
%-------------------------------------------------------
We fit to the data up to 50~MeV above the $D_s\bar D^{*0}$ threshold, i.e., 4.03~GeV using MINUIT~\cite{James:1975dr,iminuit,iminuit.jl}. 
There are only four parameters: 
$C^{(O)}(\Lambda)$, $D^{(O)}(\Lambda)$, 
an overall normalization factor,  and a relative coupling strength for diagrams (d, e) in comparison with diagrams (a, b, c) in Fig.~\ref{fig:feyn}. { Details of the fits, not given below, can be found in the Supplemental.} 

For the constant-contact case (fit I), we find
\begin{eqnarray}
  C^{(O)}(\Lambda) = -0.77^{+0.12}_{-0.10}\,\left(-0.45^{+0.05}_{-0.04}\right)\,{\rm fm}^2 \, ,
  \label{eq:fit-virtual}
\end{eqnarray}
with $\chi^2 / {\rm dof} = 0.60 (0.61)$ for $\Lambda = 0.5 (1.0)\,{\rm GeV}$; the $T$-matrix has a virtual state $Z_{cs}$ close to threshold.
For the resonant EFT (fit II), we determine
\begin{eqnarray}
  C^{(O)}(\Lambda) &=&
  -0.72^{+0.18}_{-0.13} \, \left(-0.44^{+0.06}_{-0.05}\right)\, {\rm fm}^2 \, , \notag
  \\
  D^{(O)}(\Lambda) &=& -0.17^{+0.21}_{-0.21}\,
  \left(-0.025^{+0.066}_{-0.049}\right)\,
  {\rm fm}^4 \, ,\label{eq:fit-res-d}
\end{eqnarray}
with $\chi^2 / {\rm dof} = 0.60 (0.61)$, and consequently the $Z_{cs}$ could be a resonance.
The pole position of the $Z_{cs}$ and its spin and SU(3)-flavor partners, using these parameters, are summarized in the lower half of Table~\ref{tab:predictions}.
Although the central values of the poles differ from those using the $Z_c(3900)$ inputs, they agree within uncertainties. { The largest discrepancies are found for  $C^{(O)}(\Lambda)$ in the resonant EFT [Eqs.~\eqref{eq:cd_zc} and \eqref{eq:fit-res-d}], for which variations exceed two sigmas. Nevertheless, the overall picture is qualitatively consistent with small SU(3) light-flavor corrections, although the large errors prevent us from reaching quantitative conclusions.  }

A comparison of these fits with the data is shown in Fig.~\ref{fig:fit_b} for
$\sqrt{s}=4.681$~GeV. The two cases, constant-contact and resonant EFT, can both fit the data well. This is similar to the case of the $Z_c(3900)$ in the analysis of Ref.~\cite{Albaladejo:2015lob}. Therefore, to distinguish the two scenarios, further experimental exploration with more statistics would be helpful.
The comparison for the other four energy points is shown in Fig.~\ref{fig:fit_b_4E} with the resonant EFT fit. We can see that the fit describes the five recoil-mass spectra well simultaneously.

{ Finally, we point out that statistically acceptable fits to the BESIII invariant mass distributions can be obtained after setting the parameter $r$ in the amplitudes to zero. That is, neglecting the contributions of diagrams (d) and (e) of Fig. ~\ref{fig:feyn}. The new best-fit $\chi^2/$dof and parameter errors are only slightly higher and smaller, respectively, with values for the merit function of around 0.7 now.  In addition, the $C^{(O)}(\Lambda)$ and $D^{(O)}(\Lambda)$ LECs  are little affected, with changes included in errors, while the general normalization parameter increases by 30--40\%. These readjustments are sufficient to describe the current BESIII data, and therefore it is difficult to unravel the importance of mechanisms (d)  and (e) given the available statistics.
\begin{figure}
    \centering
    \includegraphics[width=\linewidth]{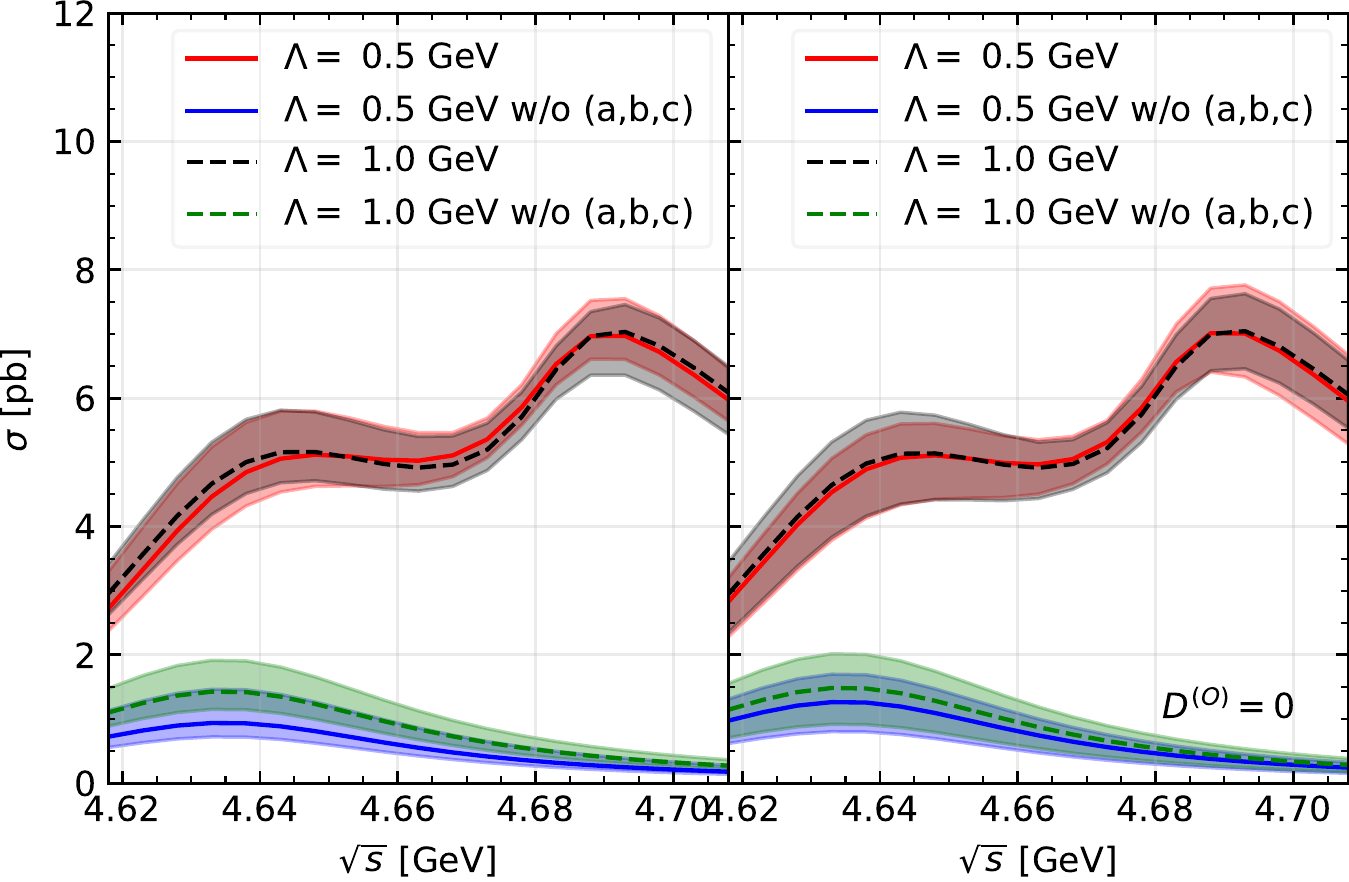}
    \caption{Predicted cross section for the $e^+e^-\to K^+(D_s^-D^{*0}+D_s^{*-}D^0)$ reaction, with the invariant mass of the open-charm meson pair restricted to to be less than 4.03~GeV, as a function of the $e^+e^-$ c.m. energy. Confident-level (68\%) bands from the correlated errors of the best-fit parameters are also displayed. The results obtained without including the mechanisms driven by the $D_{s2}$ resonance,  diagrams (a), (b) and (c)  in Fig.~\ref{fig:feyn}, are also shown (lower sets of cross section). }
    \label{fig:xs}
\end{figure}
Nonetheless, we show in Fig.~\ref{fig:xs},  the predicted $e^+e^-\to K^+(D_s^-D^{*0}+D_s^{*-}D^0)$ cross section, with the $D_s^{(*)-}D^{(*)0}$ invariant mass integrated from the threshold up to 4.03~GeV (upper limit in the fits). The cross section shows a fairly mild dependence on the $e^+e^-$ c.m. energy when the mechanisms of the diagrams (a), (b) and (c)  in Fig.~\ref{fig:feyn} 
% (exchange of the $D_{s2}$ and the subsequent triangle accounting for the final state interaction)  
are not included. Moreover, in this case we do not observe any structure above $\sqrt{s}=$4.66~GeV.  The peak around 4.69~GeV clearly indicates the importance of the (a), (b) and (c) diagrams. Unfortunately, it is unclear how the prediction of Fig.~\ref{fig:xs}  can be compared with the Born cross section $\sigma(e^+e^-\to K^+ Z_{cs}^-)\times \mathcal{B}(Z_{cs}^-\to D_s^-D^{*0}+D_s^{*-}D^0)$ reported by BESIII, since  the latter was extracted from a resonance fit, after subtracting a (model dependent) non-resonant contribution.}

{\it{Summary and outlook.}}---We have investigated the newly observed charged hidden-charm
state $Z_{cs}(3985)$ by BESIII in the processes $e^+e^-\to K^+(D^-_sD^{*0}+D^{*-}_s D^0)$. We present the first theoretical fit to all energy points, for which we consider two different EFTs describing the $Z_{cs}$. Three findings are worth noticing: First, the mass of this state does not necessarily coincide with that from the BW parametrization, and the $Z_{cs}$ could be either a virtual state or a resonance. 
Second, the near-threshold signal is further enhanced when the $e^+e^-$ energy is close to the $D_{s2}\bar D_s^*$ threshold at 4.681~GeV.
Third, the $Z_{cs}$ is probably the SU(3)-flavor partner of the previously known $Z_c(3900)$, which also implies the existence of a so far unobserved $Z_{cs}^*$ state as its spin partner.

{When only the constant interaction is considered, the generated poles are purely molecular~\cite{Guo:2017jvc,Gamermann:2009uq}. Non-molecular components enter only if energy-dependent potentials, the $D^{(O)}(\Lambda)$ term in resonant EFT, are involved.
Intuitively, this can be understood as the coupling of a different state, whatever it is, with the $Z_{cs}$ will bring a non-molecular component into the wave function. 
Yet, the data can be well fitted with only the $C^{(O)}(\Lambda)$ term and in most of the cases, the numerical value of the $D^{(O)}(\Lambda)k^2$ piece of the interaction is smaller than the $C^{(O)}(\Lambda)$ one, which may be regarded as a support of the hadronic molecular interpretation.}

The poles presented here correspond to an isospin triplet $(Z_c^\pm, Z_c^0)$ and two isospin doublets ($Z_{cs}^+,Z_{cs}^0$ and their antiparticles), as well as their spin partners. In the whole SU(3)-flavor multiplet family containing a singlet and an octet, there should be two more isospin scalars. The physical isoscalar states should be mixtures of the octet isoscalar and the singlet, the prediction of which requires more inputs.

Future experiments would be required to establish the SU(3) and spin multiplets.
High statistics data for the $e^+e^-\to J/\psi K\bar K$ at similar energies are desirable.

\medskip

{{\it Note added.}---After the submission of this paper, the LHCb Collaboration announced the observation of the $Z_{cs}(4000)$, with a mass consistent with the $Z_{cs}(3985)$ but a much larger width, and reported another $Z_{cs}(4220)$~\cite{Aaij:2021ivw}. However, it is worthwhile to notice that the LHCb fit does not describe the $J/\psi K$ distribution well around 4.1~GeV, where there is a hint of a dip. It may well be produced by the  $D^*\bar D_s^*$ molecule predicted here (see Table~\ref{tab:predictions}), as a result of its interference with coupled channels~\cite{Dong:2020hxe}.
At last, let us comment on the Argand diagram of the $Z_{cs}(4000)$ also reported in Ref.~\cite{Aaij:2021ivw}, which contains eight data points from about 3.84 to 4.17~GeV. The first four points rise quickly with little curvature, then the Argand plot turns drastically counter-clockwise to the fifth point. Such behavior is in line with the $Z_{cs}$ being a hadronic molecule coupled to a lower $J/\psi K$ channel. Indeed, the $\bar D_s^*D$ threshold occurs between the third and fourth points, and its strong coupling to the $Z_{cs}$ will lead to a cusp behavior in the Argand diagram.
}

\bigskip

\begin{acknowledgments}

This work is partly supported by the National Natural Science Foundation of China (NSFC) under Grants No.~11735003, No.~11835015, No.~12047503, No.~11975041, No.~U2032109, and No. 11961141012, by the NSFC and the Deutsche Forschungsgemeinschaft (DFG, German Research
Foundation) through the funds provided to the Sino-German Collaborative
Research Center ``Symmetries and the Emergence of Structure in QCD"
(NSFC Grant No. 12070131001, DFG Project-ID 196253076 - TRR110), by the Fundamental Research Funds for the Central Universities, by the Chinese Academy of Sciences (CAS) under Grants No.~XDB34030303 and No.~QYZDB-SSW-SYS013, by the CAS Center for
Excellence in Particle Physics (CCEPP), by the Spanish Ministerio de Economía y Competitividad (MINECO) and the European Regional
Development Fund (ERDF) under contract FIS2017-84038-C2-1-P, by the EU Horizon
2020 research and innovation programme, STRONG-2020 project, under grant agreement No.~824093,  by Generalitat
Valenciana under contract PROMETEO/2020/023, and by the CAS President’s International Fellowship Initiative (PIFI) under Grant No.~2020VMA0024.

\end{acknowledgments}

\begin{appendix}
  \section{Appendix: Fit details}

\begin{widetext}

The $\bar D_{s}^*D^0 + \bar D_{s}D^{*0}$ invariant mass distribution is given by
\begin{align}
    \frac{d\Gamma}{dm_{23}} =& \frac{{\cal N} q_K p_3^*}{\sqrt{s}}  
    \bigg\{\frac12\int_{-1}^1d\cos\theta_3^* 
    \left| q_K^2 \left[ \frac{2m_{D_{s2}}}{m_{13}^2 - m_{D_{s2}}^2 + i m_{D_{s2}} \Gamma_{D_{s2}}} + I(q_K) T_D \right] 
    + r \left[ 1 + (I_{0,\bar D_{s}^* D^0} T_D - I_{0,\bar D_{s} D^{*0}} T_E) \right] \right|^2 \notag\\
    & + \left| q_K^2I(q_K) T_E - r \left[ 1 + (I_{0,\bar D_{s}^* D^0} T_D - I_{0,\bar D_{s} D^{*0}} T_E) \right] \right|^2
    \bigg\},
    \label{eq:amp_triangle_bubble}
\end{align}
\end{widetext}
where ${\cal N} = \frac{ m_{D_s^*}m_{D^0}}{18\pi^3} g^2$ is an overall constant, $T_D$ and $T_E$ are the $T$-matrix elements for $\bar D_s^* D^0\to \bar D_s^*D^0$ and $\bar D_s^* D^0\to \bar D_s D^{*0}$, respectively, $r$ is a parameter describing the relative weight between diagrams (d,e) and diagrams (a,b,c) in Fig.~1 of the main text, $I_0$ is the two-point nonrelativistic loop integral given by Eq.~(15) in the main text, and the scalar 3-point loop integral is given by~\cite{Guo:2010ak,Guo:2017jvc,Guo:2020oqk}
\begin{widetext}
\begin{align}
  I = \frac{\mu_{12}\mu_{23}}{2\pi\sqrt{a}} \left[
\arctan\left(\frac{c_2-c_1}{2\sqrt{a(c_1-i\epsilon)}}\right) 
- \arctan\left(\frac{c_2-c_1-2a}{2\sqrt{a(c_2-a-i\epsilon)}}\right)
\right],
    \label{eq:Iexp}
\end{align}
\end{widetext}
where $\mu_{12}$ and $\mu_{23}$ are the reduced masses of the $D_{s2}\bar D^*_s$ and $\bar D_s^*D^0$, respectively,
$a = \left(\mu_{23}q_K/m_{D^0}\right)^2$, $c_1= 2\mu_{12}b_{12}$,
$c_2=2\mu_{23}b_{23}+q_K^2\mu_{23}/m_{D^0}$ with $b_{12} = m_{D_{s2}}+m_{\bar D_s^*}-\sqrt{s}$ and 
$b_{23}=m_{\bar D_s^*}+m_{D^0}+E_K-\sqrt{s}$, and $q_K (E_K)$ is the $K^+$ momentum (energy) in the $e^+e^-$ c.m. frame. 
The involved kinematic variables are given by
\begin{align}
    q_K & = \frac1{2M}\sqrt{\lambda(s,m_K^2, m_{23}^2)}, \notag\\
    m_{13}^2 &= m_K^2 + m_{D^0}^2 + 2 E_1^* E_3^* - 2 p_1^* p_3^* \cos\theta_3^*, \notag\\
    p_1^* &= \sqrt{E_1^{*2} - m_K^2},\quad p_3^* = \frac1{2m_{23}}\sqrt{\lambda(m_{23}^2, m_2^2, m_3^2)}, \notag\\
    E_3^* & = \frac{m_{23}^2 - m_2^2 + m_3^2}{2m_{23}}, \quad E_1^* = \frac{s - m_{23}^2 - m_K^2}{2m_{23}}, 
    \quad .
\end{align}
For the process with the $K^+\bar D_{s}^*D^0$ final state, $m_2 = m_{D_{s}^*}$ and $m_3 = m_{D^0}$; 
for that with the $K^+\bar D_{s}D^{*0}$ final state, $m_2 = m_{D_{s}}$ and $m_3 = m_{D^{*0}}$. To a very good approximation, $p_3^*$ for the two processes can be taken to be the same, leading to the above expression. With the single-channel approximation, we also have $T_D = - T_E = \tau(E_{\rm cm})/2$ with $\tau(E_{\rm cm})$ given by Eq.~(14) in the main text.

In the fit, we assume that the parameter $r$ is the same for each energy
point. While for the production of the process $e^+e^-\to K^+(D^-_sD^{*0} + D^{*-}_s D^0)$, we have 
\begin{equation}
    \frac{dN}{dm_{23}} = h\;\left|\frac{2m_{\psi}}{s - m_{\psi}^2 + i m_{\psi} \Gamma_{\psi}} \right|^2\frac{d\Gamma}{dm_{23}}\;\mathcal{L}_{int}\;\bar{\epsilon}\;f_\text{corr}
\end{equation}
where $\psi$ is the charmonium $\psi(4660)$. $\mathcal{L}_{int}$, $\bar{\epsilon}$ and $f_\text{corr}$
are the integrated luminosity, detection efficiency and correction factor, respectively, which
can be found in the Table~I of BESIII paper~\cite{Ablikim:2020hsk}. The factor $h$ is associated with the $e^+e^-$
annihilation vertex, here we take a same value at the energy range from 4.628 to 4.698~GeV. Thus, we
introduce the fit parameter $\tilde{\cal{N}} = {\cal{N}} h$, which is independent of c.m.
$e^+e^-$ energy. 

In the constant-contact EFT fit, we have 3 free parameters, $\tilde{\cal{N}}$, $C^{(O)}$ and $r$. 
The besfit results for $\tilde{\cal{N}}$ and $r$ are
\begin{eqnarray}
  \tilde{\cal{N}} &=&
  (0.78^{+0.16}_{-0.15})\times 10^{-3} \, \left((0.81^{+0.17}_{-0.16})\times 10^{-3}\right) \, , \notag
  \\
  r &=& -1.7^{+0.7}_{-0.8}\,
  \left(-1.2^{+0.4}_{-0.5}\right) \, ,
\end{eqnarray}
for $\Lambda = 0.5 (1.0)\,{\rm GeV}$.
The correlation matrices for $\Lambda = 0.5$~GeV and $\Lambda = 1$~GeV are
\begin{align}
    \begin{pmatrix}
   1.0 &	0.15 &	0.81 \\
	0.15 &	1.0 &	-0.28 \\
	0.81 &	-0.28 &	1.0
    \end{pmatrix}
\end{align}
and
\begin{align}
    \begin{pmatrix}
    1.0	& 0.1 &	0.82 \\
	0.1 &	1.0 &	-0.34 \\
	0.82 &	-0.34 &	1.0
    \end{pmatrix},
\end{align}
respectively. On the other hand in the resonant EFT fit, we have 4 free parameters, $\tilde{\cal{N}}$, 
$C^{(O)}$, $D^{(O)}$ and $r$. The fitted $\tilde{\cal{N}}$ and $r$ parameters are
\begin{eqnarray}
  \tilde{\cal{N}} &=&
  (0.79^{+0.16}_{-0.15})\times 10^{-3} \, \left((0.81^{+0.17}_{-0.16})\times 10^{-3}\right) \, , \notag
  \\
  r &=& -1.9^{+0.6}_{-0.7}\,
  \left(-1.3^{+0.4}_{-0.5}\right) \, ,
\end{eqnarray}
for $\Lambda = 0.5 (1.0)\,{\rm GeV}$. The correlation matrix for $\Lambda = 0.5$~GeV is
\begin{align}
    \begin{pmatrix}
      1.0 &  0    & -0.04 & 0.85 \\
       0 &  1.0    & -0.58 & -0.2 \\
   -0.04 & -0.58 &  1.0    & -0.13 \\
    0.85 & -0.2  & -0.13 & 1.0
    \end{pmatrix}
\end{align}
while for $\Lambda = 1$~GeV is
\begin{align}
    \begin{pmatrix}
    1.0 &	0.09 &	-0.07 &	0.85 \\
	0.09 &	1.0 &	-0.67 &	-0.16 \\
	-0.07 &	-0.67 &	1.0 &	-0.07 \\
	0.85 &	-0.16 &	-0.07 &	1.0
    \end{pmatrix}.
\end{align}

\end{appendix}

\bibliography{Zc-SU3.bib}
 
\end{document}